\documentclass[aps,prfluids]{revtex4-2}
\usepackage[utf8]{inputenc} % allow utf-8 input
\usepackage[T1]{fontenc}    % use 8-bit T1 fonts
\usepackage{hyperref}       % hyperlinks
\usepackage{booktabs}       % professional-quality tables
\usepackage{amsfonts}       % blackboard math symbols
\usepackage{nicefrac}       % compact symbols for 1/2, etc.
\usepackage{microtype}      % microtypography
\usepackage{lipsum}
\usepackage{graphicx}
\usepackage{amsmath}
\usepackage{textgreek}
\begin{document}

\preprint{APS/123-QED}

\title{Dynamics of Freely Suspended Drops Translating through Miscible Environments}

\author{Endre Joachim Mossige} 
 \thanks{Equal contribution}
% \email{endre.mossige@gmail.com}
  \affiliation{
 Department of Chemical Engineering, Stanford University, Stanford, California 94305
} 
 
\author{Vinny Chandran Suja}
%\email{vinny@stanford.edu} 
\thanks{Equal contribution}
\affiliation{
 Department of Chemical Engineering, Stanford University, Stanford, California 94305
}

\author{Daniel J. Walls}
\thanks{Equal contribution}
%\email{daniel.joseph.walls@gmail.com}
\affiliation{
 Department of Chemical Engineering, Stanford University, Stanford, California 94305
}
  
\author{Gerald G Fuller}
\email{ggf@stanford.edu} 
\affiliation{
 Department of Chemical Engineering, Stanford University, Stanford, California 94305
}

  %% \AND
  %% Coauthor \\
  %% Affiliation \\
  %% Address \\
  %% \texttt{email} \\
  %% \And
  %% Coauthor \\
  %% Affiliation \\
  %% Address \\
  %% \texttt{email} \\
  %% \And
  %% Coauthor \\
  %% Affiliation \\
  %% Address \\
  %% \texttt{email} \\

\begin{abstract}

%This chapter focuses on an experimental investigation of droplets freely rising through a miscible, more viscous liquid. We report observations of droplets of water rising through glycerol and corn syrup. By varying the initial drop size and the ambient liquid, the relative importance of convective to diffusive transport mechanisms is manipulated. This balance is captured by the Rayleigh number, which in our experiments span from $0.58\cdot10^4$ (for the smallest water drops suspended in corn syrup) to $210\cdot10^4$ (for the largest drops suspended in glycerol). As droplets rise, their volume increases and their velocity decreases, eventually following power laws at long times. The drops continually grow more oblate as they rise. (Second paper: Interestingly, we estimate the interfacial tension between water and glycerol by applying the developments of Taylor and Acrivos \cite{taylor1964} to our experiments.)

Our work focuses on an experimental investigation of droplets freely rising through a miscible, more viscous liquid. We report observations of water droplets rising through glycerol and corn syrup, which are common household ingredients. Immediately after the drops are formed, they take on prolate shapes and rise with constant velocity without expanding in size. However, after a critical time predicted by our theory, the drops continually grow into oblate spheroids, and as they mix with the ambient liquid, their volume increases and their velocity decreases, eventually following power laws. We present scaling relations that explain the main observed phenomena. However, the power laws governing the rate of the volumetric increase and the velocity decrease, namely $t^{1/2}$ and $t^{-1/2}$, respectively, remain points of further investigation.

\end{abstract}
\maketitle
\section{Introduction}
\label{sec:intro}
% keywords can be removed

Miscible liquids often come into contact with one another in natural and technological situations, as well as in the kitchen, commonly as a drop of one liquid present in a second miscible liquid. Examples include the uptake of an injectable drug in the blood stream and the mixing of a drop of freshwater with seawater when the river meets the sea. Familiar examples from the kitchen include the dissolution of honey in tea and the mixing of dish washing liquid with water. This paper focuses on droplets freely suspended in a second, miscible liquid that is unbounded and quiescent. Specifically, the droplet is formed from a liquid of lower density and lower viscosity than the miscible, ambient liquid in which it is suspended, such that it rises due to buoyancy forces. To date, miscible configurations of freely suspended droplets have been restricted to studies where a droplet of greater density and viscosity impacts upon an air-liquid interface, and then continues to descend into the second liquid of lower density and viscosity. Significantly, inverting the direction of the buoyancy force allows an inversion of the viscosity ratio to be examined, as within sets of miscible liquids, the denser liquid tends to be more viscous as well. 

The translation of droplets and bubbles through a viscous liquid is a classic problem in fluid mechanics. Hadamard \cite{hadamard1911} and Rybczynski \cite{rybczynski1911} each studied spherical droplets translating through an immiscible liquid at low Reynolds number. Hadamard and Rybczynski developed, independent of one another, the solution for the stream functions inside and outside the droplet.

Stokes' drag law describes the drag on a solid sphere translating through a viscous fluid at low Reynolds number \cite{stokes1851effect2,dey2019terminal}. The terminal velocity is reached when buoyancy is exactly balanced by the Stokes' drag, and is given by 

\begin{equation}
    u_0=\frac{2a^2\Delta\rho g}{9\mu},
    \label{eq:terminalVelocity}
\end{equation}
where $a$ is the sphere radius, $\Delta \rho$ is the density difference, and $g$ the acceleration due to gravity. For a droplet or bubble rising or falling in another fluid, the  viscous stress is continuous across the fluid-fluid interface, leading to a mobile fluid-fluid interface. The degree of interface mobility is determined by the viscosity ratio between the inner and outer liquid, $\lambda\equiv\hat{\mu}/\mu$, and the magnitude of this parameter modifies the drag force. In the limit $\lambda \rightarrow \infty$, the interface is completely immobile, causing the drop velocity to match that of a buoyant, rigid sphere. However, 
in the opposite limit, $\lambda \rightarrow 0$, the skin friction drag vanishes, causing an increase in the terminal velocity by 50 $\%$. 

%The streamlines for the streamfunction are displayed in Figure \ref{fig:hadamardrybczynski}.

%\begin{figure}[!h] \centering \includegraphics[width=8cm]{img/hadamardrybczynski.pdf}
% \begin{figure}[!h] \centering \includegraphics[width=8cm]{fig1.pdf} \caption[ Streamlines of the Hadamard-Rybczynski solution.]{Streamlines of the Hadamard-Rybczynski solution given by Equations (\ref{eq:outerpsi}) and (\ref{eq:innerpsi}). \label{fig:hadamardrybczynski}} \end{figure}

%Taylor \& Acrivos \cite{taylor1964} later revisited the problem of spherical drops to consider deviations from an exactly spherical shape.
A number of studies have investigated influences of drop shape on the dynamics of translating drops. The first consideration of deviations from the exactly spherical shape was carried out by Taylor \& Acrivos  \cite{taylor1964}, who investigated influences of finite inertia on the shape evolution of falling drops. They determined that finite inertia causes the drop to deform into an oblate spheroid for moderate values of the Weber number, $We=\Delta\rho au_0^2/\sigma$, where $\sigma$ is the interfacial tension between the droplet and surrounding fluid. As $We$ increases, the droplet deforms further into a spherical cap. Several years later, Kojima \emph{et al.} \cite{kojima1984} reported a study of drops of aqueous corn syrup solutions falling into bodies of water and then descending; they coupled their experiments with theoretical work, tracking the shape evolution of the droplets as they descended within the bath of water. In their analytical treatment, they look at the evolution of the descending drop for two limiting cases during its descent: just after submersion and at longer times after the formation of an open toroidal structure. They found that finite and time-dependent interfacial tensions as well as small, but finite contributions of inertia both contributed to the shape evolution of miscible, falling drops. However, they did not incorporate effects of diffusion in their analysis, and they especially note the neglect of an expanding diffuse interface as the result of dissolution. 
%(Interestingly, Taylor \& Acrivos demonstrate the insensitivity of the deformation to the ratio of viscosities between the two liquids, $\lambda$. As the shape of a droplet deviates from a sphere, the drag it experiences changes as well. Taylor \& Acrivos incorporate this aspect into their study, referring to the work of Payne \& Pell \cite{payne1960}, which investigates axially symmetric bodies in Stokes flow. Harper \cite{harper1972} provides a substantive review of bubbles and droplets translating through various mediums.)

Arecchi \emph{et al.} \cite{arecchi1989} reported an experimental study of droplets of mixtures of glycerol and water impacting an air-water interface at high velocities, corresponding to a high $We$ number configuration. In some experiments, they observed droplets to break up upon descending within the bath of water; whereas in others, they observed no break up. They also reported a dimensionless number that compares the time scales of convective mixing and diffusion along with an empirical analysis for a critical value of this parameter that determines whether fragmentation occurs or not. In a followup study, Arecchi \emph{et al.} \cite{arecchi1996} extended their prior study by considering the particular features of the fragmentation process. Almost twenty years later, Walker, Logia and Fuller \cite{walker2015multiphase} utilized developments in high speed photography to investigate glycerol drops impacting on water and found the dynamics to be highly dependent upon the transient interfacial tension between these miscible liquids. 
%(The stabilizing effect of ultra-low tensions between miscible liquids is thoroughly described in the book by Joseph and Renardy \cite{joseph1993}.)

Another body of research concerns the influence of the initial condition on the drop shape at long times. 
Koh \& Leal \cite{koh1989,koh1990} studied the stability of droplets of silicone oils rising through an immiscible, viscous castor oil at low Reynolds number A clever experimental apparatus including a plunger allowed for the distortion of the droplet into an oblate or prolate spheroid as an initial condition. The authors found that after small perturbations, the droplet could recover to a spherical shape, whereas large perturbations resulted in instabilities including cases where unstable prolate spheroids developed tails that pinched off due to capillary forces, and unstable oblate spheroids developed skirts that likewise pinched off due to the same mechanism.

% [VINNY: I LIKE THE REST OF THE INTRO, TELL ME WHAT YOU THINK! I will keep working on the paragraphs leading up to this. - I agree. For the four paragraphs immediately prior to this - can you combine/reorganize them so that the paragraphs focus on the research and not on individual studies?]
Thermal plumes, or thermals, are created by heating a fluid from below, and represent yet another class of miscible fluids. In a study investigating the implications of heated plumes emanating from volcanoes on the aviation industry, G.I. Taylor \cite{taylor1946dynamics} was the first to investigate the dynamics of turbulent plumes as early as 1946. The transport of both mass and momentum for turbulent plumes is dominated by advection rather than diffusion, leading to high Reynolds and Rayleigh numbers. In 1960, Morton \cite{morton1960weak} performed a theoretical study of extremely weak thermals, where transport of heat and momentum are both dominated by diffusion, leading to both small Rayleigh and Reynolds numbers. Morton points out that such weak thermals are difficult to produce in a laboratory due to the restriction on the thermal blob size. Interestingly, the height and the diameter of thermal structures are reported to follow a square root dependence on time in both of these two opposite limits. However, the height risen by the thermal also depends on the Rayleigh number as defined by the initial condition of the thermal blob, and this dependence differs drastically between different types of thermals. 
%For example, the height of turbulent plumes are reported to depend on the Rayleigh number to the power of one fourth, while diffusion dominated thermals depend more strongly on this parameter, yielding a linear dependence.  

A third regime is identified for extremely viscous and buoyant drops rising in colder, more viscous environments, where advection dominates the transport of heat, but not the transport of momentum, leading to large Rayleigh numbers, but small Reynolds numbers. Griffiths \cite{griffiths1986thermals} found these types of thermals to expand as they rise, however as a consequence of the large Rayleigh number, the volumetric growth was not found to be  attributed to diffusion alone across the miscible interface. Instead, Griffiths found viscous thermals to continually expand due to convective flows as they entrain ambient liquid during their rise, leading to a volumetric growth dependence of time as $V\sim t^{3/2}$. The thermals maintained a spherical shape, and this led him to argue that the viscous drag stayed relatively constant. As a result, he concluded that the decaying velocity of the thermals, $U\sim t^{-1/2}$, was merely a consequence of their increasing volume. 
%A third regime is identified for extremely viscous and buoyant drops rising in colder, more viscous environments and this is the focus of the report by Griffiths \cite{griffiths1986thermals}, who studied heated blobs of extremely viscous artificial polybutene oils rising in the same liquid.

Diffusion acts to smear out the interface between a freely suspended drop and its environment, however this is also the fate of sessile and pendant drops submerged in miscible liquids. In two separate reports from this laboratory, we have previously studied the shape evolution of relatively dense and viscous drops submerged in a lighter, less viscous ambient liquid \cite{walls2016spreading,walls2018shape}. Initially, the drops swell as diffusion across the miscible interface proceeds, however beyond a critical thickness of the evolving miscible layer, gravitational forces act to deform this layer, thereby draining the interfacial drop material. For sessile drops, the interfacial material was found to spread radially outward as a viscous gravity current \cite{huppert1982}, while for the pendant drops the interfacial layers flow out in the form of a jet from the drop's apex. %The speed of the advancing front follows $r\sim t^{1/2}$, which is slightly faster than the advancing contact line at the substrate, $r\sim t^{1/8}$. For pendant drops, the interfacial smearing mechanisms causes a jet to form at the apex of the drop, causing the drop volume to decrease exponentially in time. %The onset time of the jetting flow was predicted as $\tau \sim a^2Pe^{-2/3}/D_0$, where $a$ is the drop radius, $Pe$ the Peclet number ($Pe=U_0a/D_0$), with $u_0$ being a characteristic velocity, and $D_0$ the mutual diffusion coefficient between the pure substances. The above scaling was found to efficiently predict a similar transition between convection and diffusion dominated behavior for the rising miscible drops considered in this paper.

%By smearing out the interface between the dense drop material and the ambient liquid over time, the interfacial viscosity drops dramatically, and this facilitates weak flows driven by gravity into the ambient phase.  In the case of sessile drops, the act of gravity leads to a viscous gravity current, which spreads as $\sim$XX, leading to the drop volume shrinking as $\sim$XX. For pendant drops, we report a jetting phenomenon..., leading to a volumetric decline  of XX"]
%He found the heated liquid stayed contained within the drops, which led the buoyancy force being constant as well.

%From Griffiths (1986): "Viscous thermals are also of interest in the field of chemical engineering, where the shape of, and heat and mass transfer from, buoyant drops has been widely investigated. However, it appears that exchange by molecular diffusion between the drop and its surroundings has been discussed only for the case of immiscible (spherical) drops (Kronigt $\&$ Brink 1951;Pan & Acrivos 1968;Brignell 1975), while the behaviour of miscible drops in Stokes flow, coupled with diffusion of buoyancy, has not been considered beyond a single interesting observation by Kojima, Hinch & Acrivos (1984)."

%Since the drops translate at low Reynolds number, their dynamics can be treated as quasi-steady. This means that the viscous drag is relatively constant on the time scale describing the dynamics of thermals. 

In this paper, we report the dynamics of droplets freely rising in relatively more viscous and more dense miscible liquids. Specifically we focus on  evaluating the evolution of the shape, size and velocity of these drops at low Reynolds and high Peclet number. For this purpose, experiments were performed with water drops rising in glycerol and corn syrup. In section \ref{sec:experimental}, we detail the experimental setup and protocols used in this study. Subsequently, in section \ref{sec:theory} we develop a theoretical framework describing the relevant timescales in the system. The key findings from this study are reported and discussed in section \ref{sec:Results}, and we conclude the manuscript in section V.

%In order to best explain the experimental results, we start by developing a theoretical understanding of rising miscible drops. To assist in this development, we build on our physical intuition. 

%----------------------------------------------------------
\section{Experimental}\label{sec:experimental}

\subsection{Miscible Rising Drop Apparatus}

\begin{figure}[!h] \centering \includegraphics[width=\linewidth]{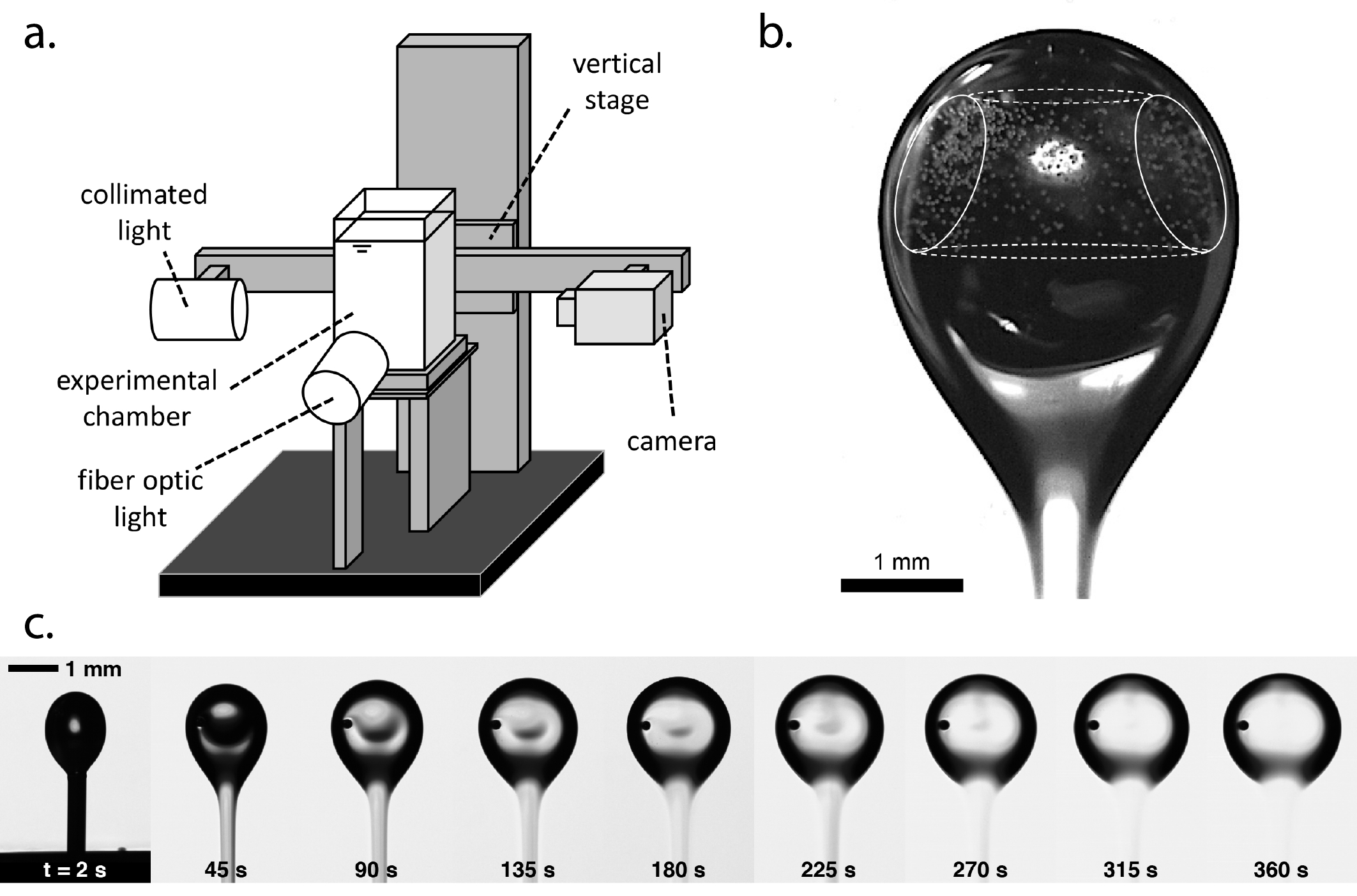} 
\caption[]{(a) Diagram of the experimental setup for performing miscible rising drop experiments. First, the experimental chamber is filled with the ambient liquid. Then, the rising drop is formed in the ambient liquid by using a syringe needle inserted through a hole in the aluminum base of the chamber. Once the droplet is formed, it begins to rise immediately, and the needle is withdrawn once the droplet has risen several radii above its tip. A camera is mounted on a vertical stage and positioned to image the rising droplet from the side. A fiber optic light is mounted orthogonal to the camera to allow for particle tracking. (b) As a result of the flow field developing within the drop as it rises, the tracer particles used for flow visualization form a toroidal ring structure. (c) Image sequence of a water drop rising in glycerol. The drop takes on a prolate shape when formed by the needle (t=2s) and transitions to an oblate spheroid while growing in size as it translates through the denser, more viscous ambient liquid.  \label{fig:risingexp}} 
\end{figure}

An apparatus was developed to perform measurements of the evolution of droplets of a less viscous and less dense liquid rising through a more viscous and more dense, miscible ambient liquid. The apparatus consists of an experimental chamber, constructed from a base of aluminum and four walls formed by four glass microscope slides. The dimensions of the chamber are 50 mm (l) x 50 mm (w) x 150 mm (h). A narrow vertical hole is drilled through the center of the aluminum base and is sealed with a silicone film. To perform an experiment, the chamber is first filled with the ambient liquid, and the droplet of the rising liquid is formed by piercing the seal with a syringe and injecting the rising material. Once the droplet is formed, it begins to rise immediately. Once it has risen several radii above the tip of the syringe needle, the needle is withdrawn from the chamber. A camera (Model: GPF 125C IRF, Allied Vision Technologies, PA, USA) is positioned to view the droplet from the side, and is mounted on a vertical translation stage (Model: ULM-TILT, Newport, CA, USA) to track the drop as it rises. The camera is fitted with a telecentric lens (Model: 63074, Edmund Optics, NJ, USA), and is backlit by a fiber optic light (Model: 21AC fiber optic illuminator, Edmund Optics, NJ, USA) that passes through a collimator (Model: 62760, Edmund Optics, NJ, USA)  in order to best distinguish the liquid-liquid interface between the miscible liquids. The apparatus and experimental setup are both  displayed schematically in Figure \ref{fig:risingexp}a.

Rising drops of deionized water (Milli-Q Academic A10) were formed in  glycerol (Fisher; Certified ACS) and corn syrup (Karo; light corn syrup). The densities and viscosities of these liquids are displayed in Table \ref{tab:liqprop}. The initial volume of rising droplets was varied between 1 \textmu l and 10 \textmu l. A stainless steel needle with an outer diameter of 0.362 mm and a syringe (10 \textmu l, Hamilton Microliter$^{\mathrm{TM}}$~\#801) were used to form the rising droplet in the experimental chamber.

\begin{table}[h]
\centering
\begin{tabular}{|l c c c c c c c c|}
\hline
Liquid & {} & Rising & {} & Density & {} & Viscosity & {} & Diffusion coefficient \\
{} & {} & or ambient & {} & (g/mL) & {} & (mPa$\cdot$s) & {} & ($10^{-4}$mm$^2$/s)\\
\hline
Water & {} & Rising & {} & 0.997 & {} & 0.890 & {} & - \\
Glycerol & {} & Ambient & {} & 1.261\cite{glycerine1963physical} & {} & 925 & {} & 0.9 \\
Corn Syrup & {} & Ambient & {} & 1.386 & {} & 5100 & {} & 1.3\cite{ray2007} \\
\hline
\end{tabular}
\caption{Densities ($\rho$) and viscosities ($\mu$) of the experimental liquids, and mutual diffusion coefficients ($D_0$) between pairs of liquids. Standard cone plate rheometry was employed to obtain the viscosities.} 
\label{tab:liqprop} 
\end{table}

\subsection{Particle Tracking Velocimetry}

A fiber optic cable was positioned to project light through the side of the experimental chamber orthogonal to the camera. Microsphere particles were added to the rising liquid droplet to scatter the incident light, allowing the particles to be imaged. All particle imaging experiments were performed with the rising liquid containing 35 $\mu$m-diameter microspheres at a concentration of $10^{-2}$ g/ml (Cospheric; UVPMS-BY2, 32-38 $\mu$m fluorescent yellow polyethylene microspheres, $\rho_{particle}=1.0$ g/cm$^3$) and sodium dodecyl sulfate (Invitrogen; UltraPure$^{\mathrm{TM}}$) at a concentration of $10^{-3}$ g/ml to disperse the microspheres  by adsorbing onto their surfaces. Figure \ref{fig:risingexp}b is a representative image showing particles dispersed in a droplet as it rises. The particles form a ring around the drop perimeter, and this observation is explained in the Results section below. 

In order to best explain the experimental results, we start by developing a theoretical understanding of rising miscible drops. 
%To assist in this development, we build on our physical intuition. 

\section{Theory}\label{sec:theory}

\subsection*{Dominating mechanisms determining the dynamics of miscible rising drops }
\label{sec:theory}

As soon as a water drop is formed, it starts to mix with the surrounding liquid by molecular diffusion. The time scale describing this process depends on the molecular diffusion coefficient between the fluids, $D_0$, and the drop radius, $a$, and is given by 

\begin{equation}
    \tau_{diff}\sim a^2/D_0.
    \label{eq:diffusionTimeScale}
\end{equation} 

Diffusion acts on a slow time scale, as can be seen by considering a typical $\text{1\,mm}$ sized water drop submerged in glycerol, for which $\tau_{diff}\sim 10^4\text{\,s}$. 

However, a recirculating flow pattern develops within the drop as it rises, as seen in Figure 1b, and this flow field causes the concentration gradient between the outer liquid and the drop liquid to be restricted to a thin miscible layer, which is the characteristic length scale for the present mixing problem. The extent of this interfacial layer, $l$, was found by Acrivos and Goddard \cite{acrivos1965} for the analogous transport problem concerning heat transfer from a sphere in a uniform flow at low Reynolds number and high Péclet number ($Pe$) to be:   %and these flows accelerate the mixing process through so-called Taylor dispersion \cite{taylor1953dispersion,aris1956dispersion,frankel1989foundations,bryden1999mass}, leading to an enhanced "effective" diffusivity, $D_{eff.}$. $D_{eff.}$ depends on $D_0$, the Pecl$\acute{e}$t number, $Pe=u_0a/D_0$, and a geometric constant, $K$, and is given as

% \begin{equation}
%     D_{eff.}=D_0\left(1+KPe^2\right).
%     \label{eq:Deff}
% \end{equation}
% For $Pe\gg\text{1}$, mass is primarily transported by convection, while for small values, mass is primarily transported by diffusion. For the translating miscible drops treated in this paper, $Pe\sim10^4$ $\gg\text{1}$, so that $D_{eff.}\sim D_0 Pe^2$.

% The convective flow also 

\begin{equation}
    l \sim aPe^{-1/3}
    \label{eq:AcrivosGoddard1965},
\end{equation}
where $Pe$ depends on $a$, $D_0$ and the characteristic velocity, $u_0$, and is defined as $Pe=u_0a/D_0$.
In order to arrive at the correct time scale of the present mass transfer problem concerning rising miscible drops, we substitute $l$ for $a$ in Equation \ref{eq:diffusionTimeScale} to obtain a new time scale that incorporates effects of convection:

\begin{equation}
    \tau \sim \frac{a^2 Pe^{-2/3}}{D_0} = \tau_{diff}Pe^{-2/3}.
    \label{eq:tau}
\end{equation}
Note that the above time scale is faster than the diffusion time scale by a factor of $Pe^{-2/3}$. For the problem at hand, this difference is considerable as $Pe\sim10^4$, resulting in a mixing time that is typically 100 times less than that of a pure diffusion process. 

The buoyant drops considered in this paper have no prescribed velocity, but instead the terminal rise velocity $u_0$ is defined by the Stokes' velocity in  Equation \ref{eq:terminalVelocity}. Incorporating $u_0$ into the definition of the Péclet number, we obtain the well known Rayleigh number. Utilizing this information in Eq. \ref{eq:tau} and simplifying, we obtain a characteristic time scale for rising miscible drops, which is a typical transition time between diffusion-free and diffusion-influenced behavior:

\begin{equation}
    \tau \sim \left(\frac{\mu^2}{\left(\Delta\rho g\right)^2 D_0}\right)^{1/3}.
    \label{eq:diffusionConvectionTimescale}
\end{equation}
We will use this time scale to rationalize the onset of the velocity and volumetric change of miscible drops rising in viscous environments.

\section{Results and Discussion}\label{sec:Results}

\subsection{Qualitative observations} 
\label{subsec:shapeEvolution}
%%%%%%% width = 17.6cm     or     height=1cm
% \begin{figure}[!h] \centering %\includegraphics[width=\textwidth]{img/risingsequence.pdf}
% \includegraphics[width=\textwidth]{fig1.pdf}
% \caption[Brightfield image sequence taken in time of a droplet of water freely rising in glycerol.]{Brightfield image sequence taken in time of a droplet of water freely rising in glycerol. In the first frame, the droplet has just been injected and sits atop the syringe needle. As the droplet rises, it increases in volume due to diffusion and a trailing strand of diffuse liquid is shed at its rear stagnation point. Additionally, the droplet begins to deviate from a spherical shape, becoming more oblate as it rises. For the full movie from which these frames were taken, contact the author. \label{fig:risingsequence}} \end{figure}

Figure \ref{fig:risingexp}c shows a representative series of images taken of a droplet of water freely rising in glycerol. In the first frame, the droplet has just been injected and sits atop the syringe needle. The droplet takes on a prolate shape, and this initial shape was consistent across all of the drops in our experiments. However, as the drop rises, it becomes more spherical. Throughout the rise, diffusion acts to smear out the interface, leading to a diminishing interfacial tension. This leads to a reduction in the capillary pressure jump across the drop, making the droplet more susceptible to interfacial deformations. As a result, the droplet deviates from the spherical shape at long times,  becoming increasingly more oblate.
%The observed transition from prolate to spherical is expected for small deviations as explained by Koh \& Leal \cite{koh1990}.

At the apex of the droplet, the interface between the droplet and ambient liquid remains distinct throughout the experiment. As we move towards the rear stagnation region the drop interface grows fainter, where it sheds a trailing strand of diffuse liquid, as seen in Figure \ref{fig:risingexp}b and c. The liquid shed by the drop is initially produced when the drop first rises above the syringe needle, indicating that it is at least a partial consequence of the technique for producing the droplet. However, this strand is continually produced as some of the material in the diffuse layer at the liquid-liquid interface is swept behind the droplet as it rises. The strand neither stretches behind the droplet nor breaks up due to capillary forces as was reported by Koh \& Leal \cite{koh1990} for immiscible rising drops; in time, the strand continues to diffuse into the ambient liquid. The absence of capillary breakup of the strand into droplets is a consequence of the low interfacial tension between the material forming the strand and the ambient liquid. 

As is evident from Figure \ref{fig:risingexp}c, the volume of the drop increases with distance traveled. This observation is attributed to effects of diffusion, however, the drop volume does not increase solely through diffusion of external fluid into the drop as in the low Péclet number limit \cite{morton1960weak}. Instead, the volumetric growth is a result of both diffusion and convection; as miscible liquid in the layer forming around the drop is dragged to the rear stagnation point by viscous stresses, the strong recirculating flow carries this miscible fluid to the interior of the drop. As explained by Griffiths \cite{griffiths1986thermals} for the analogous case of viscous thermals at low Re and high Pe, the miscible fluid approaching the rear stagnation region is most likely to rise with the droplet rather than to be pulled away by the trailing strand, since it is less dense than the ambient fluid.

%(As the droplet of water rises, diffusion across the miscible liquid-liquid boundary proceeds.) 

%As seen in Figure \ref{fig:risingexp}c, the boundary appears distinct towards the apex of the drop and begins to grow fainter towards the rear stagnation point. 

%Figure \ref{fig:risingsequenceinitial} shows a series of images taken of the same experiment in Figure \ref{fig:risingsequence} during the initial seconds of the droplet rising above the syringe needle.
%Initially, all of our drops take on a prolate shape when formed by the needle. However, the frontal part of the drop becomes more spherical as it rises in the ambient liquid and this observation is in agreement with results of  immiscible rising drops reported previously by Koh \& Leal \cite{koh1989}. 

%Additionally, the drop begins to deviate from a spherical shape, becoming more oblate as it rises. 

In Particle Tracking Velocimetry (PTV) experiments, particles were observed to concentrate and circulate in an open toroidal structure within the drop. Similar observations were reported by Griffiths \cite{griffiths1986thermals} for rising thermals. A representative image from these experiments is shown in Figure \ref{fig:risingexp}b. As the particles are restricted to certain portions of the droplet, it is difficult to surmise the structure of the flow in the droplet where particles are not observed. It is expected that similarly structured paths to the streamlines of Hadamard \cite{hadamard1911} and Rybczynski \cite{rybczynski1911} exist, although slightly modified due to the distortion of the droplet from a spherical shape.

%As described in a previous paper by our group  treating the shape evolution of pendant and sessile droplets being both more dense and more viscous than the ambient phase (Walls \emph{et al.} \cite{walls2018shape}), the dynamics of miscible drops is slow enough such that a diffuse momentum boundary layer evolves at the miscible fluid-fluid interface. In our previous experimental configuration with pendant and sessile drops, the boundary layer could be observed directly, as it manifested within the ambient liquid of lower viscosity and density. But in the present experiments with rising droplets, the viscosities of the phases are inverted, resulting in the diffuse boundary layer growing \emph{inside} of the droplet; thus, it is obscured from view, as light passing through the edge of the droplet is deflected away from the camera. Future studies may use of fluorescence imaging, which has proven to be useful in revealing details in situations where the curvature of liquid-liquid interfaces obscure regions of a brightfield image. 

The following sections aim to give a more quantitative description of the observed phenomena, and we start by describing the shape evolution.  

\subsection{Shape evolution of rising miscible drops}

Throughout the experiment, we track the shape of the rising drop. We identify the boundary of the top hemisphere of the rising drop as a distinct transition in optical contrast. The boundary of the bottom hemisphere of the droplet is defined to exclude the trailing strand, which is done by using the distinct boundary of the upper half of the  hemisphere to fit a parabola in order to enclose the lower part of the droplet. The volume, the centroid, and the major and minor axes of this boundary is measured as a function of time.

\begin{figure}[!h] \centering %\includegraphics[width=8cm]{img/risingoblatedrop.pdf}
\includegraphics[width=\linewidth]{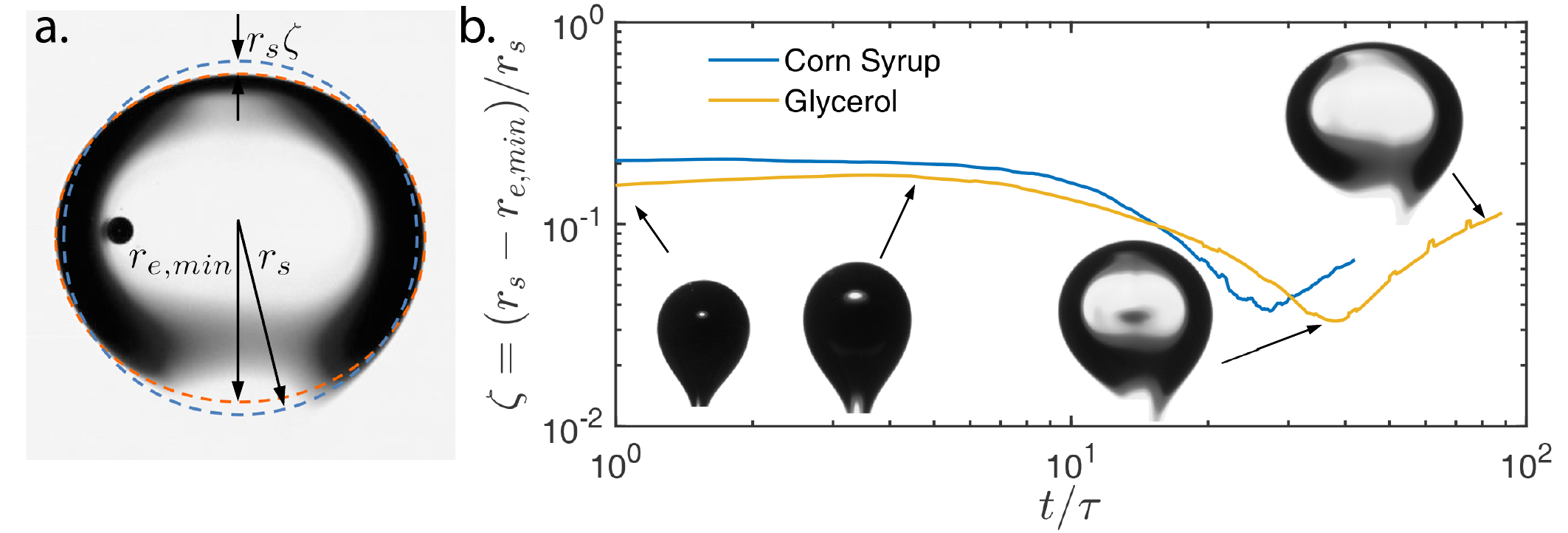}
\caption{{\bf a.} Image of a rising drop of water in glycerol that has deformed into an oblate spheroid. The dashed orange oval traces the visible boundary of the oblate rising droplet. The dashed blue circle indicates the boundary of a sphere with an identical volume to the oblate rising droplet. The radial deformation of the droplet from a spherical shape to an oblate spheroid is indicated by $\zeta$. Here $\zeta = (r_s - r_{e,min})/r_s$, $r_s$ the radius of equivalent spherical drop and $r_{e,min}$ the minor axis of the actual spheroid. {\bf b.} Evolution of $\zeta$ for water drops in corn syrup and glycerol, showing a prolate to oblate transition in shape. As a consequence of the higher diffusion coefficient in the glycerol-water system as compared to the corn syrup-water system, drops rising through glycerol deform more than drops rising through corn syrup.   }\label{fig:risingoblate} \end{figure}

In order to describe the interfacial deformation of the droplet from a prolate to an oblate spheroid, we introduce the dimensionless coordinate, $\zeta = (r_s - r_{e,min})/r_s$, that depends on the radius of an equivalent spherical drop, $r_s$, and the minor axis of the actual spheroid, $r_{e,min}$, see Figure \ref{fig:risingoblate}a. In Figure \ref{fig:risingoblate}b, $\zeta$ is plotted against dimensionless time, $t/\tau$. The coordinate $\zeta$ stays constant for some time after the drop is released from the needle, which is consistent with the observation of the drop maintaining a prolate shape. After a critical time of $t\sim\tau$, $\zeta$ begins to decrease and reaches a minimum as the drop transitions towards a spherical shape. After the minimal value is reached, $\zeta$ starts to increase as the drop develops into a highly deformed oblate spheroid. 

The degree of drop deformation is governed by a competition between the shear stresses acting on the drop from the ambient liquid and a small capillary pressure across the drop interface. As soon as the drop is formed, molecular diffusion acts to smear out the interface, lowering the transient tension and making it susceptible to interfacial deformations. After some time, $t\sim\tau$, the interface is sufficiently smeared that ambient liquid is allowed to enter the drop through its rear. This onset of convective mixing further diminishes the transient tension, which in turn accelerates the rate at which the interface is deformed, and as the tension continues to diminish, the drop becomes increasingly oblate, as seen by the increasing value of $\zeta$ in Figure 2b. Drops rising in glycerol end up as being more oblate than drops rising in corn syrup, and this is due to the relatively large value of the diffusion coefficient in the glycerol-water system (see Table I), reducing the structural integrity of these drops. 

Miscible drops descending in another miscible liquid was reported by Kojima \emph{et al.} \cite{kojima1984} to  evolve from an oblate spheroid into an open torus. In our experiments, this evolution was not observed, as the drop continually became more oblate as it rises, but never deviated from this basic shape. The difference between their observations and ours can be explained by the (exactly) reversed viscosity ratio; our study concerns water drops rising through corn syrup, while their study concerns drops of corn syrup descending through water. As reported by Kojima \emph{et al.}, the evolution from oblate drops to toroidal shapes is attributed to inertial effects, which can be safely ignored in our experiments due to the high viscosity of the ambient phase as compared to the drop phase. By applying a syringe and needle, a simple kitchen flow experiment can be performed to confirm the different behavior reported for the two different fluid-fluid systems described here, and we encourage the reader to do so.  

\subsection{Evolution of drop volume and velocity}
\label{subsec:DropVolumeAndVelocity}

Throughout the experiment, the drop volume, $V$, increases, and the velocity, $u$, decreases. In order to quantify these observations, we track $V$ and $u$ of  rising drops. $V$ is calculated assuming axial symmetry about the vertical axis, and $u$ is calculated by measuring the displacement of its center of mass between successive frames and dividing by the time elapsed. We start our description by considering the evolution of the drop velocity, $u$.

%INTRO? As we will shortly see, $u$ and $V$ are closely inter-connected. 

\subsubsection{Evolution of drop velocity}
\label{subsec:DropVelocity}

\begin{figure}[hp] \centering %\includegraphics[width=\textwidth]{img/risingvelocity.pdf}
\includegraphics[width=\textwidth]{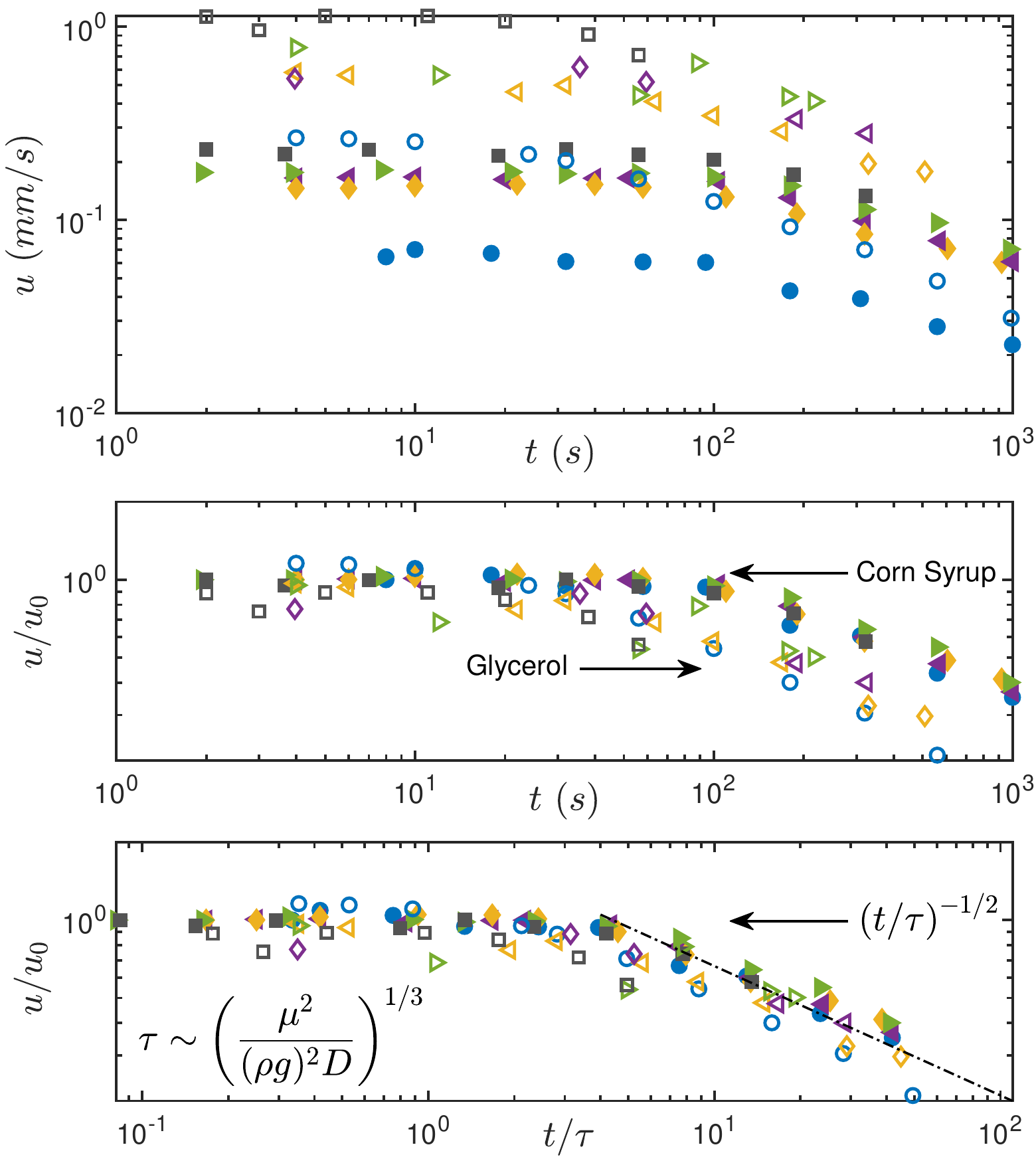} \caption[Plots of velocities as functions of time for rising drops of water in glycerol and corn syrup.]{Plots of velocities as functions of time for rising drops of water in glycerol and corn syrup. The data is plotted on logarithmic scales. {\bf a.} Unscaled velocity data. {\bf b.} Velocity is scaled by the initial velocity of each rising drop, $u_0$, and time is left unscaled. Varying symbol shapes and colors represent discrete experiments. Open and filled symbols indicate the ambient liquid as glycerol and corn syrup, respectively. {\bf c.} Velocity is scaled by $u_0$, and time is scaled with a viscous time scale $\tau \sim \left(\frac{\mu^2}{(\rho g)^2D} \right)^{1/3}$. The dashed black line represents a power law of $t^{-1/2}$. \label{fig:risingvelocity}} \end{figure}

Figure \ref{fig:risingvelocity}a shows a plot of the time dependent velocities of water drops rising in glycerol and in corn syrup for a range of initial drop volumes. The data is plotted on logarithmic scales. Initially, the velocities of the rising droplets remain constant, before beginning to diminish, ultimately following a power law at long times. The decrease in velocity is expected, as the droplet increases its drag as it expands.

If we make the velocities presented in Figure \ref{fig:risingvelocity}a non-dimensional by the initial velocity, $u_0$, the plot shown in Figure \ref{fig:risingvelocity}b is obtained. Scaling the velocity by the initial velocity of each rising drop collapses the vertical axis, and allows us to best observe the behavior of the velocity at long times. This scaling is sufficient to obtain a single response of the data generated for a particular ambient liquid, but not across different ambient liquids.

Figure \ref{fig:risingvelocity}c shows the resulting velocity data when time is made dimensionless by the viscous time scale, $\tau$. For small values of $t/\tau$, the velocities stay relatively constant, while for larger values of this parameter, the velocities of droplets rising in both ambient liquids appear to asymptote to a power law of $t^{-1/2}$. The slightly steeper decline observed for glycerol drops is most likely due to the relatively larger deformation, as can be seen from Figure \ref{fig:risingoblate}, giving rise to a larger drag force for a given drop volume. 

For $t/\tau\textless1$, mass transfer is dominated by convection, with effects of diffusion being negligible, resulting in the drops being prolate or spherical with the buoyancy force exactly balancing the viscous drag force, leading to constant (terminal) velocities. As dimensionless time proceeds, however, effects of diffusion can no longer be ignored, as it causes the drops to expand, thereby increasing the drag experienced by the drops. The buoyancy force is expected to stay relatively constant during the rise as the buoyant fluid is mostly contained within the drop without escaping through diffusion. As argued by Griffiths for the analogous system of rising thermals \cite{griffiths1986thermals}, this behavior is a consequence of the high Péclet number, since mass does not have time to diffuse across the miscible interface on the time scale of the flow. Therefore, the observed velocity decrease at long times is most likely dominated by the increasing viscous drag caused by the increasing volume. 
The power law of $-1/2$ should be predicted by the power laws governing the evolution of the volume, density, and viscosity of the droplet through the balance of buoyancy and drag forces, but unfortunately, the scaling parameters cannot be used to fully explain this behavior. 

%Note that there is a slight discrepancy between the observed initial velocities, $u_0$, and the theoretical velocities calculated from the Hadamard-Rybzynski solution, $u_{0,HR}$. For all droplets, a $20\%$ smaller than expected initial velocity is observed, and for the smallest drops (< 2 $\mu$l), the velocity ratio decreases further. Similar behavior across the two ambient liquids is observed. For bubbles and drops rising or falling through viscous liquids, it has been observed that below a certain volume ($\sim$1 $\mu$l), the bubble rises at a velocity that is two-thirds that of the Hadamard-Rybczynski solution \cite{leal2007}. It has been suggested that the reduction in the velocity is due to impurities that contaminate the air-liquid or liquid-liquid interface, essentially acting as a surfactant and creating a Marangoni stress that hinders the rise of the bubble or droplet \cite{ervik2017transition}. This behavior is also reported for most "pure" liquids without any added surfactants, such as water drops rising through castor oil \cite{silvey1916fall,bond1927lxxxii,bond1928lxxxii}, and suggests that only trace amounts of impurities is sufficient to immobilize the interface, causing the observed velocity reduction. Therefore, we infer that Marangoni stresses due to contaminants adsorbing to the liquid-liquid interface are likely to be the cause of the reduced velocity of miscible drops reported in this paper. 

\subsubsection{Evolution of drop volume}
\label{subsec:DropVolume}

\begin{figure}[hp] \centering %\includegraphics[width=\textwidth]{img/risingvolume.pdf}
\includegraphics[width=\linewidth]{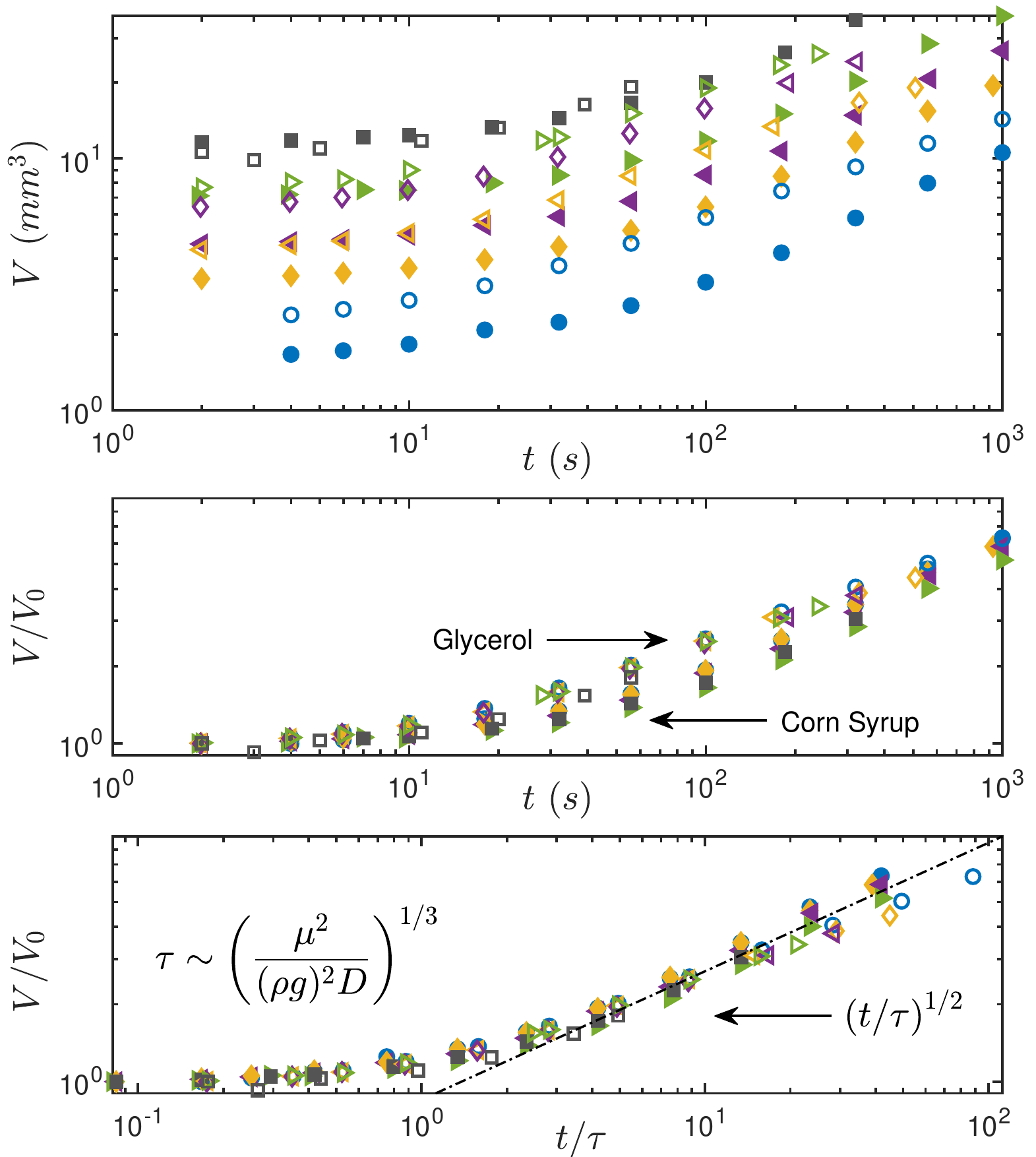}
\caption[Plots of volumetric measurements as functions of time for rising drops of water in glycerol and corn syrup.]{Plots of volumetric measurements as functions of time for rising drops of water in glycerol and corn syrup. The data are plotted on logarithmic scales. {\bf a.} Unscaled data. {\bf b.} The volumetric data are scaled by the initial volume of each rising drop, $V_o$, and time is left unscaled. Varying symbol shapes and colors represent discrete experiments. Open and filled symbols indicate the ambient liquid as glycerol and corn syrup, respectively. {\bf c.} Time scaled with a time scale $\tau \sim \left(\frac{\mu^2}{(\rho g)^2D} \right)^{1/3}$. The dashed black line represents a power law of $t^{1/2}$. \label{fig:risingvolume}} \end{figure}

Figure \ref{fig:risingvolume}a shows a plot of the volumetric measurements as a function of time for the rising drops of water in glycerol and corn syrup for a range of initial volumes. The data is plotted on logarithmic scales. Initially, the volumes of the rising droplets increase only slightly before growing as a power law at longer times. The increase in volume occurs as ambient liquid is swept across the liquid-liquid interface into the interior of the droplet. Simultaneously, as the droplet rises, the interior liquid circulates and flows over this diffuse layer, acting to convect the diffuse material away from the layer, thus promoting additional diffusion.

%Throughout the experiment, the drop volume, $V$, increases as ambient fluid in swept into the drops during its rise. 

If we scale the volumetric data presented in Figure \ref{fig:risingvolume}a by the initial volume, $V_o$, and leave the time-axis unscaled, the plot shown in Figure \ref{fig:risingvolume}b is obtained. Scaling the volume by the initial volume of each rising drop collapses the vertical axis ($\bar{V}=V/V_o$) for each liquid system, but not across the two different systems.  

The plot in Figure \ref{fig:risingvolume}c shows the response of the volumetric data scaled by the initial drop volumes and time scaled by the viscous time scale, $\tau$. The scaling of time successfully collapses the two data sets, however there is some spread at long times. The data set demonstrates the appropriateness of the viscous time scale, $\tau$, for predicting the onset of diffusion on the volumetric growth of miscible drops. For $t/\tau$<1, the drops display immiscible behavior, as seen by the near constant values of the scaled volumes, while for $t/\tau$>1, the drop volumes start to increase. The latter behavior is a a result of the interface being more and more diffuse as the drop rises, allowing ambient liquid to enter the interior of the drop leading to the volumetric growth.

An interesting observation is the volumetric growth rate with the exponent of $1/2$ with respect to dimensionless time. Unfortunately, our scaling analysis is not able to explain the observed power law. However, it should be noted that a diffusive process would indicate a power law of $3/2$ and we cannot rule out the possibility that convection is influencing the volumetric growth rate.

%Another point of investigation concerns the observed power law of volumetric growth. A power law of $t^{1/2}$ indicates a diffusive process for a one-dimensional length, but is not expected for a three-dimensional quantity such as volume. Further, it is not entirely clear that droplets rising through the two different ambient liquids asymptote exactly to the same power law.

\subsection{Influence of the trailing strand}
In the analysis, we have ignored influences of the trailing strand on the shape and velocity evolution of translating drops. This might seem like a over simplification at first glance. However, our data shows qualitative agreements with data previously reported for translating  viscous drops, where trailing strands are absent. For example, in the early stages, the shape evolution of the miscible drops reported here agree reasonably well with the immiscible drops reported by Koh \& Leal \cite{koh1990}. Later in the rise, our drops develop into oblate spheroids, and this observation is in agreement with descending miscible drops reported previously by Kojima, Hinch and Acrivos  \cite{kojima1984}, where trailing strands are absent. 

In order for the trailing strand to severely impact the dynamics of rising miscible drops, the mass loss throughout the rise must be significant, even early in the rise where the drops display "immiscible" behavior. However, our data shows that the mass within the drop stays relatively constant in the early stages of the rise, and that the mass loss is negligible. Later in the rise, the mass starts to increase due to influences of diffusion. Another evidence of the negligible influence of the trailing strand is the constant velocity maintained at early times, which indicates that the buoyancy force remains constant. The diminishing velocity at long times is a manifestation of diffusion, which causes the drop to grow bigger as more and more ambient liquid is swept into the drop from the back as it rises. Finally, Griffiths argues that the miscible boundary layer forming around the drops characterized by low Re and high Ra (Pe) as described here tends to rise with the drop rather than to a pulled away to form a trailing strand, and this is a consequence of the buoyancy force largely exceeding the viscous dissipation of momentum. We expect the drops described in this paper to behave qualitatively similar to the rising thermal reported by Griffiths. 
%Griffiths argues that the heated liquid stayed contained within the drop as a consequence of the buoyancy force largely exceeding the viscous dissipation of momentum

The qualitative agreement between the shape evolution and velocity of rising droplets reported here with drops reported previously led us to ignore the trailing strand in the theoretical description. However, our theory was not able to fully explain the observed power law behavior for volume, and we cannot rule out the possibility of the trailing strand as being, at least in part, the cause of this discrepancy. %between theory and experiment.

%Griffiths: large Ra. Cannot use the observed rising velocity for $t/\tau$<1 to explain the influence of the trailing strand. 

%This power law should be predicted by the power laws governing the evolution of the volume, density, and viscosity of the droplet through the balance of buoyancy and drag forces. Likewise, the relation of the power laws observed for volume and velocity should predict the power law for density, or the concentration, $c$, of the drop. Relating these power laws through this force balance,
%\begin{equation} \frac{4}{3}\pi a^3g\Delta\rho=4\pi a\mu U\left[\frac{3\lambda+2}{2\left(\lambda+1\right)}\right], \label{eq:dragbuoyancy} \end{equation}
%predicts that the density, or concentration, of the rising liquid should diminish like $t^{-11/15}$ as it rises at long times. This power law indicates that the droplet does not only incorporate ambient liquid into its interior as it rises, but that it also sheds liquid from the initial volume as well, which is supported by the visual observations in Figure \ref{fig:risingsequence} of the trailing strand.

\section{Conclusion}\label{sec:conclusion
}
To the best of our knowledge, this is the first experimental study of droplets freely rising in more viscous and more dense, miscible liquids. We demonstrate that droplets rising through miscible environments increase in volume with their velocity decreasing, and we postulate this behavior as being a consequence of ambient liquid being swept into the drops as they rise. Our theoretical study revealed the correct time scale and length scale to explain key observations. Particularly, we are able to predict the transition from immiscible behavior, where diffusion can be ignored, to a diffusion influenced regime, where the drops develop into oblate spheroids, with the velocity decreasing and the volume increasing.  However, our scaling parameters are insufficient in explaining the rate of the volumetric expansion and the velocity reduction at long times, $t^{1/2}$ and $t^{-1/2}$, respectively, which remain points of further investigation.  

%Future work: Future studies may use of fluorescence imaging, which has proven to be useful in revealing details in situations where the curvature of liquid-liquid interfaces obscure regions of a brightfield image. 

%As noted, additional experiments and theoretical investigations should reveal the correct time scale to explain the observations.

%To evaluate the predicted behavior of the concentration within the droplet, a clever experiment using food coloring could be devised to correlate the diminishing pigment in the drop to the decreasing concentration.

%The slow character of the evolution of these rising droplets is encouraging in using such an experiment to estimate the time-dependence of the interfacial tension between miscible liquids.

%Experiments in a taller experimental chamber could be conducted to see if an open torus develops at very long times, or if the drop simply continues to grow more oblate.

\section{Acknowledgements}
We would like to thank Yidi Tai for her assistance in performing experiments with rising drops and Matt Chuck for his guidance in fabricating pieces of the experimental apparatus.
%I would like to acknowledge the support of my adviser, Prof. Gerald G. Fuller, in allowing the exploration of this configuration of miscible liquids. 

\section{Data Availability}
The data that support the findings of this study are available from the corresponding author upon reasonable request.

\bibliography{References}  %%% Remove comment to use the external .bib file (using bibtex).

%\begin{thebibliography}{1}

%\bibitem{kour2014real}
%George Kour and Raid Saabne.
%\newblock Real-time segmentation of on-line handwritten arabic script.
%\newblock In {\em Frontiers in Handwriting Recognition (ICFHR), 2014 14th
%  International Conference on}, pages 417--422. IEEE, 2014.

%\bibitem{kour2014fast}
%George Kour and Raid Saabne.
%\newblock Fast classification of handwritten on-line arabic characters.
%\newblock In {\em Soft Computing and Pattern Recognition (SoCPaR), 2014 6th
%  International Conference of}, pages 312--318. IEEE, 2014.

%\bibitem{hadash2018estimate}
%Guy Hadash, Einat Kermany, Boaz Carmeli, Ofer Lavi, George Kour, and Alon
%  Jacovi.
%\newblock Estimate and replace: A novel approach to integrating deep neural
%  networks with existing applications.
%newblock {\em arXiv preprint arXiv:1804.09028}, 2018.

%\end{thebibliography}

\end{document}